\documentclass[aps,12pt,final,notitlepage,oneside,nobibnotes,nofootinbib,superscriptaddress,onecolumn,noshowpacs,amsmath,amssymb]{revtex4}

\usepackage[english]{babel}
\usepackage[utf8x]{inputenc}
\usepackage[T1]{fontenc}

\usepackage{amsmath}
\usepackage{graphicx}
\usepackage[colorinlistoftodos]{todonotes}
\usepackage[colorlinks=true, allcolors=blue]{hyperref}
\usepackage{url}

\newcommand{\nder}{\!\!\!\not\partial}

\begin{document}

\title{Chiral Condensate in Two Dimensional Models}
\author{V.G. Ksenzov}
\affiliation{State Scientific Center Institute for Theoretical and Experimental Physics, Moscow, 117218, Russia}

\begin{abstract}
We investigate  two different models. In one of them  massive fermions interact with a massive scalar field and in the other the fermion field is in an electrical field (QED2). The chiral condensates are calculated in one-loop approximation. We found that the chiral condensate in the case of the Yukawa interaction the fermions and scalar field does not vanish if the mass of the fermion field tends to zero. The chiral condensate disappears in QED2, if the fermion mass is zero. 
\end{abstract}

\maketitle

\section{Introduction}

Dynamical breaking of the symmetry is described by a quantity known as the order parameter. In models with a broken  chiral symmetry the chiral condensate is the order parameter that disappears, if the chiral symmetry is restored. The investigation of the chiral condensate plays a crucial role in attempts to describe phase transitions related to the dynamical chiral symmetry breaking.

In QCD chiral condensate is investigated via lattice simulations. This approach is usually employed in studies of the chiral condensate in the presence of external factors such as temperature, chemical potential, magnetic fields and magnetic monopoles, etc (see Refs.~\cite{BCL}--\cite{RS} and references therein).

Apart from numerical studies there are a lot of models employed for the investigation of the chiral condensate in various physical systems \cite{BBM}--\cite{JLK}, \cite{NJL}--\cite{KR2}. In the first paper on this subject Nambu and Jona-Lasinio (NJL) analyzed a specific field model in four dimensions \cite{NJL}. Later,  the chiral condensate was studied by Gross and Neveu (GN) in two dimensions spacetime in the limit of a large number of fermion flavors $N$ \cite{GN}. These two models are similar, but in contrast to NJL-model, GN-model is a renormalizable and asymptotically free theory. Due to these properties, GN-model is used for qualitative modeling of QCD. The relative simplicity of both models is a consequence of the quartic fermion interaction.

In our previous papers we investigated a system of a self-interacting massive scalar and massless fermion fields with the Yukawa interaction in a $(1+1)$-dimensional spacetime. In the limit of a large mass of the scalar field, the model equivalent to the GN-model. The chiral condensate was obtained by the path integral using the method of the stationary phase \cite{KR}--\cite{KR2}.

In this paper we present a study of two models. One of them is the model with  massive fermions and massive scalars with the Yukawa interaction between these two fields. It is worth to note, that the scalars is not a self-interacting fields. The other model is QED in two-dimensional space-time.

The purpose of this paper is to obtain the chiral condensate in QED2-model. To do this we must obtain an effective action of the model. Evaluating the one-loop effective potential is equivalent to the summation of an infinite class of Feynman diagrams, therefore we are unable to calculate a fermionic determinant with an arbitrary vector potential $A_{\mu}(x)$. For this reason we begin studying of the first model, that we use to investigate the effective potential and the chiral condensate, if discrete chiral symmetry was explicitly broken. This model exhibits the essential feature of the techniques, that we use to construct the effective action of QED2. In particular we can see, how the chiral symmetry breaking manifests itself in the effective potential. As a result we found the chiral condensate in the electrical field.

\section{Fermions in a scalar field}

The model, that will be discussed in this section involves $N$ massive fermion fields and a massive scalar field with the Yukawa interaction between those two fields in $(1+1)$-dimensional spacetime.

Lagrangian of the model is
\begin{equation}\label{eq1}
  L=L_b+L_f=\frac{1}{2}(\partial_{\mu}\phi)^2-\frac{1}{2}\mu^2\phi^2(x)+i\bar{\psi}^a\nder\psi^a-g\left(\phi(x)-\frac{m}{g}\right)\bar{\psi}^a\psi^a,
\end{equation}
here $\phi(x)$ is a real scalar field, $\psi^a(x)$ is a fermion fields, index $a$ runs from 1 to $N\gg 1$ and $m$ is a mass of the fermions.

The model with the massless fermions was investigated in our previous papers \cite{KR}--\cite{KR2}. In the case, if $m=0$, the Lagrangian is invariant under a discrete symmetry
\begin{equation}\label{eq2}
\psi^a\to\gamma_5\psi^a,\;\bar{\psi}^a\to-\bar
{\psi}^a\gamma_5,\;\phi\to-\phi,
\end{equation}
which is broken by the chiral condensate. If $m\neq 0$, the discrete symmetry (\ref{eq2}) disappears, and chiral condensate is not a qualitative criterion.

As before in our papers we will determine the chiral condensate by an effective potential. For our purpose we formally define the chiral correlator, using a functional integral in Minkowski space
\begin{equation}\label{eq3}
  \left\langle0|g\bar{\psi}^a\psi^a|0\right\rangle=\frac{1}{Z}\int D\phi D\bar{\psi}^aD\psi^ag\bar{\psi}^a\psi^a \exp\left(i\int d^2xL(x)\right),
\end{equation}
here $Z$ is a normalization constant. The chiral correlator (\ref{eq3}) is rewritten as
\begin{equation}\label{eq4}
  \left\langle0|g\bar{\psi}^a\psi^a|0\right\rangle=\frac{1}{Z}\int D\phi \exp\left({i\int d^2xL_b(x)}\right)i\frac{\delta}{\delta\phi}\int D\bar{\psi}^aD\psi^a \exp\left(i\int d^2xL_f(x)\right).
\end{equation}
The fermionic Lagrangian is quadratic in the field and we can integrate over them, getting
$$\left\langle0|g\bar{\psi}^a\psi^a|0\right\rangle=\frac{1}{Z}\int D\phi \frac{g^2 N\left(\phi(x)-\frac{m}{g}\right)}{2\pi}\ln\frac{g^2\left(\phi(x)-\frac{m}{g}\right)^2}{\Lambda^2}\times$$
\begin{equation}\label{eq5}
  \times\exp\left(i\int d^2x\left(\frac{1}{2}(\partial_{\mu}\phi)^2-\frac{\mu^2}{2}\phi^2-\frac{N g^2\left(\phi(x)-\frac{m}{g}\right)^2}{4\pi}\left(\ln\frac{g^2\left(\phi(x)-\frac{m}{g}\right)^2}{\Lambda^2}-1\right)\right)\right),
\end{equation}
here $\Lambda$ is the ultraviolet cutoff. 

We want to obtain the chiral condensate in the framework of one-loop approximation. Therefore we calculate (\ref{eq5}) using the method of the stationary phase. A minimum of the effective action of the system is reached if the effective potential and kinetic energy are minimal on its own:
\begin{equation}\label{eq6}
\partial_{\mu}\phi=0 \text{ and } U_{\text{eff}}(\phi)=\min.
\end{equation}
Let the constant scalar field $\phi_m$ satisfies the condition (\ref{eq6}). The factor in front of the exponent in (\ref{eq5}) is fixed at the point $\phi=\phi_m$ and we obtain
\begin{equation}\label{eq7}
  \left\langle0|g\bar{\psi}^a\psi^a|0\right\rangle=\frac{Ng^2\left(\phi_m-\frac{m}{g}\right)}{2\pi}\ln\frac{g^2\left(\phi_m-\frac{m}{g}\right)^2} {\Lambda^2}.
\end{equation}
It is worth to note, that the correlator $\left\langle0|g\bar{\psi}^a\psi^a|0\right\rangle$ is a chiral condensate only in the minimum of the effective potential $\phi_m$. The effective potential and the chiral condensate require renormalization. We renormalize effective potential following Coleman and Weinberg \cite{CW} and Gross and Neveu \cite{GN} by demanding that
\begin{equation}\label{eq8}
  \left.\frac{d^2U_{\text{eff}}}{d\phi_m^2}\right|_{\phi_m=M^2}=\mu_R^2.
\end{equation}
Then the renormalized chiral condensate is written as
\begin{equation}\label{eq9}
  \left\langle0|g\bar{\psi}^a\psi^a|0\right\rangle_R=\frac{Ng^2\left(\phi_m-\frac{m}{g}\right)}{2\pi}\ln\frac{\left(\phi_m-\frac{m}{g}\right)^2}{M^2}.
\end{equation}
$\phi_m$ is determined by mean the renormalized effective potential $U^R_{\text{eff}}$, which is written as
\begin{equation}\label{eq10}
U^R_{\text{eff}}=\frac{1}{2}\mu_R^2\phi^2+\frac{g^2N}{4\pi}\left(\phi-\frac{m}{g}\right)^2\left(\ln\frac{\left(\phi-\frac{m}{g}\right)^2}{M^2}-3\right),
\end{equation}

\begin{figure}
  \centering
  \includegraphics[scale=1.4]{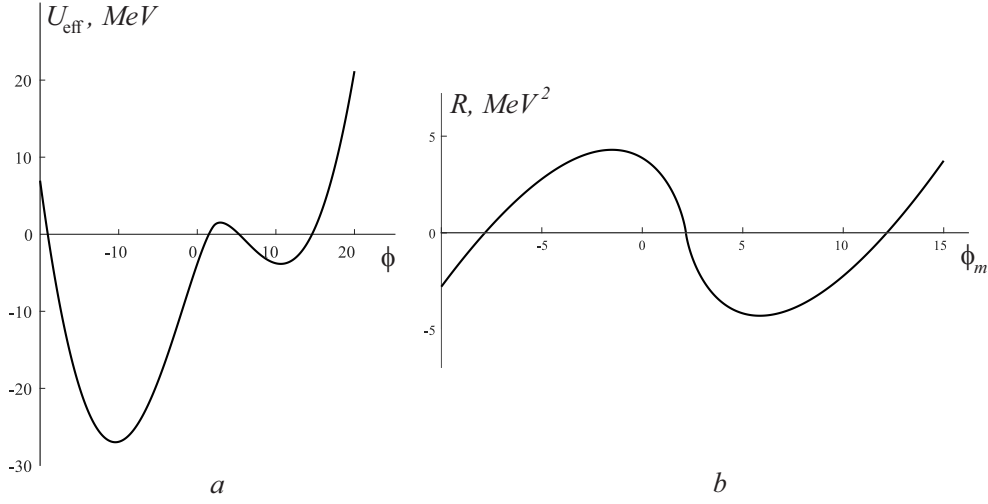}\\
  \caption{a -- The value $U_{\text{eff}}(\phi)$; b -- Chiral condensate $R=\left\langle0|m\bar{\psi}^a\psi^a|0\right\rangle_R$ at $g^2N=1,69\text{ MeV}^2$, $\mu^2_R=0,5\text{ MeV}^2$, $m/g=2,17$, $M^2=100$.}\label{Ris_1}
\end{figure}

We determine the stationary point $\phi_m$ by numerical solution of equation
\begin{equation}\label{eq11}
\left.\frac{dU_{\text{eff}}}{d\phi}\right|_{\phi=\phi_m}=\mu^2_R\phi_m+\frac{g^2N}{2\pi}\left(\phi_m-\frac{m}{g}\right)\left(\ln\frac{\left(\phi_m-\frac{m}{g}\right)^2}{M^2}-2\right)=0.
\end{equation}
One can see, that there are two different solutions of (\ref{eq11}), and for this reason appear two different vacuums. One of them is a global, and the other is a local minima (see Fig.~\ref{Ris_1}a).

Using (\ref{eq9}) and (\ref{eq11}) we get


\begin{equation}\label{eq12}
\left\langle0|m\bar{\psi}^a\psi^a|0\right\rangle_R=\frac{m}{g}\frac{N}{\pi}(g^2-g^2_{\text{cr}})\phi_m-\frac{m^2N}{\pi},
\end{equation}
here $g^2_{\text{cr}}=\frac{m^2_R\pi}{N}$.

It is worth to note, that the chiral condensate does not disappear, if the fermions mass tends to zero. In such a case $(m=0)$ the effective potential has the minimum at the point $\phi^2_m$, which is given in an explicit form
\begin{equation}\label{eq13}
\phi^2_m=M^2\exp 2\left(1-\frac{\pi \mu^2_R}{g^2N}\right),
\end{equation}
and the vacuum energy is
\begin{equation}\label{eq14}
\mathcal{E}_V=-\frac{g^2N}{4\pi}\phi^2_m.
\end{equation}
If $g^2N=\pi\mu^2_R$, then $\phi^2_m=M^2$ and the vacuum energy is completely perturbative one. It is known, that the vacuum energy is determined by the vacuum condensate of the trace of the energy-momentum tensor $\theta_{\mu\mu}$ \cite{MS}, \cite{SH}, \cite{SVZ} and \cite{SVNZ} as
\begin{equation}\label{eq15}
\frac{1}{d}\left\langle0|\theta_{\mu\mu}|0\right\rangle=\mathcal{E}_V,
\end{equation}
here $d$ is the dimension of the spacetime.

\begin{figure}
  \centering
  \includegraphics{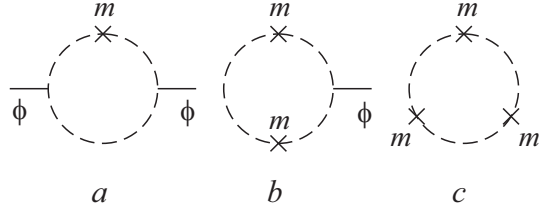}\\
  \caption{The dashed line denotes the fermion loops, solid line denotes the scalar field $\phi$.}\label{Ris_2}
\end{figure}

The diagrams shown in Fig.~\ref{Ris_2} determine the quantum correction to the trace of the energy-momentum tensor in the case of the  massive fermions.
Then the quantum correction to the trace of the energy-momentum tensor is written as
\begin{equation}\label{eq16}
\theta_{\mu\mu}=\frac{N}{2\pi}(g^2\phi^2-2mg\phi+m^2),
\end{equation}
here the fist term defines quantum anomaly if mass fermions are zero, the second term defines the discrete symmetry breaking Fig.~\ref{Ris_2}b, and the term $\propto m^2$ defines the vacuum energy noninteracting fermions Fig.~\ref{Ris_2}c.

It will be noted, that although the discrete symmetry is broken in the model with the free massive fermions, but there is no violation of this symmetry in the effective potential of the theory. This is due to the fact, that it is impossible to construct a combination of the energy dimension, that violates this symmetry.

The diagrams shown in Fig.~\ref{Ris_2} determine the quantum correction to the trace of the energy-momentum tensor in the case of massive fermions.

\section{Fermions in an electrical field}

The model that will be discussed  here is QED2.
A Lagrangian of the model is well-known
\begin{equation}\label{eq17}
  L=L_b+L_f=-\frac{1}{4}F^2_{\mu\nu}+i\bar{\psi}^a\nder\psi^a-m\bar{\psi}^a\psi^a-A_{\mu}\bar{\psi}^a\gamma_{\mu}\psi^a,
\end{equation}

If $m=0$ then the Lagrangian is invariant under a discrete chiral simmetry     
\begin{equation}\label{eq18}
\psi^a\to\gamma_5\psi^a,\;\bar{\psi}^a\to-\bar
{\psi}^a\gamma_5.
\end{equation}

We formally define the chiral correlator as
\begin{equation}\label{eq19}
  \begin{split}
      & \int d^2x\left\langle0|m\bar{\psi}^a\psi^a|0\right\rangle=\\
       =\frac{1}{Z}\int DA_{\mu} \exp\left(i\int d^2x \right. & \left. L_b(x)\!\!\!\!\!\!\phantom{\int}\right) im\frac{d}{dm}\int D\bar{\psi}^a D\psi^a \exp\left(i\int d^2x L_f(x)\right).
  \end{split}
\end{equation}

It is hard enough to  calculate a fermionic determinant with an arbitrary vector potential $A_{\mu}(x)$ therefore  we will use an alternative method of getting an effective potential.

It is known that the expression of the effective  potential was derived  within gluodynamics by Migdal and Shifman \cite{MS} and \cite{SH}. In the paper \cite{VK} the method was used to obtain the vacuum condensate of the trace of the energy-momentum tensor in massless theories in various spacetime dimensions. In this paper the method will be used for construction  the effective Lagrangian in QED2. The expression for the effective Lagrangian was obtained from the requirement that  the anomaly be reproduced under a scale transformation. The effective potential has the form
\begin{equation}\label{eq20}
V_{\text{eff}}=\frac{1}{d}\sigma\left(\ln\frac{\sigma}{\upsilon}-1\right),
\end{equation}
here $\sigma=\theta_{\mu\mu}$ and $\upsilon$ emerges as a constant of integration in solving the respective differential equation \cite{MS}, \cite{SH}. We should find $\theta_{\mu\mu}$ and $\upsilon$ for our model.

\begin{figure}[h!]
  \centering
  \includegraphics{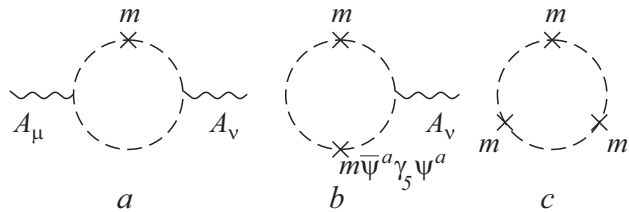}\\
  \caption{The dashed line denotes the fermion loops, solid line denotes the vector potential  $A_\mu$.}\label{Ris_3}
\end{figure}

Above we see that the mass term is fundamentally important for calculating of the chiral condensate (\ref{eq19}) therefore we must 
construct $\theta_{\mu\mu}$ taking into account the mass term.
The diagrams shown at Fig.~\ref{Ris_3} determine $\theta_{\mu\mu}$ in QED2. The chiral symmetry breaking at the expense of the mass term is given by the diagram Fig.~\ref{Ris_3}b. A result of calculating the diagrams is given
\begin{equation}\label{21}
\theta_{\mu\mu}=-e^2_{0}\frac{E^2}{6\pi}-e_{0}m\frac{E}{3\pi}+\frac{m^2}{2\pi},
\end{equation}
here we introduced a dimensionless coupling constant $e^2_{0}=\frac{e^2}{m^2}$ and an electrical field $E=\varepsilon_{\mu\nu}\partial_\mu A_{\nu}$.

To find $\upsilon$  we assume that the coupling constant $e_{0}$ is zero at the time then $V_{\text{eff}}(\sigma)$ coincides with the effective potential for free fermions and we get $\upsilon=\frac{1}{2\pi}M_{0}^2$, here $M_{0}$ is an arbitrary substraction parameter.

Now the effective Lagrangian can be expressed as
\begin{equation}\label{22}
L_{\text{eff}}=\frac{E^2}{2}+ \frac{\sigma}{2}\left(\ln\frac{2\pi\sigma}{M^2_0}-1\right).
\end{equation}
A minimum of the effective Lagrangian of the system is reached if
\begin{equation}\label{23}
\frac{d L_{\text{eff}}}{d E}=E+\frac{1}{2} \frac{d\sigma}{d E}\ln\frac{2\pi\sigma}{M^2_0}=0,
\end{equation}
here
\begin{equation}\label{24}
\frac{d\sigma}{d E}=-e^2_{0}\frac{E}{3\pi}-e_{0}\frac{m}{3\pi} 
\end{equation} 
The chiral correlation is 
\begin{equation}\label{25}
  \int d^2x\left\langle0|g\bar{\psi}^a\psi^a|0\right\rangle=\int d^2x\frac{m}{2}\frac{d\sigma}{d m}\ln\frac{2\pi \sigma}{M_{0}}, 
\end{equation}  
 here 
\begin{equation}\label{26} 
m\frac{d\sigma}{d m}=-e_{0}\frac{m E}{3\pi}+\frac{m^2}{\pi}.
\end{equation}
 
Let's $E_{m}$ is the solution of the equation (\ref{23}) then using (\ref{23}) and (\ref{25}) and accounting that $E_{m}$ and $\left\langle0|g\bar{\psi}^a\psi^a|0\right\rangle$ are constants we get
\begin{equation}
\left\langle0|g\bar{\psi}^a\psi^a|0\right\rangle=-m E_{m} \frac{d\sigma}{d m}\left(\frac{d\sigma}{d E_{m}}\right)^{-1}   
\end{equation}
or
\begin{equation}\label{27}
\left\langle0|g\bar{\psi}^a\psi^a|0\right\rangle=\left(3m^2-e_{0}m E_{m}\right)\frac{E_{m}}{e_{0} m+e^2_{0} E_{m}}
\end{equation}

In generally there are no analytical solutions (\ref{23}) but  we can find a few  of them in special cases 
if
\begin{enumerate}
    \item $m=0$, then $E_{m}=0$ and the chiral condensate disappears,
    \item $e_{0}\ll1$, then $E_{M}\simeq\cfrac {e_{0}m}{6\pi}\ln\cfrac{m^2}{M^2_{0}}$ and
$$\left\langle0|g\bar{\psi}^a\psi^a|0\right\rangle=\frac{m^2}{2\pi}\ln\frac{m^2}{M^2_{0}}\left(1-\frac{e_{0}m}{18\pi}\ln\frac{m^2}{M^2_{0}}\right),$$
here  $\left| \cfrac{e_{0}m}{6\pi}\ln\cfrac{m^2}{M^2_{0}}\right| \leqslant 1.$
\end{enumerate}

It is worth to note that $E$ may be a constant quantity. Really, fixing a gauge $A_1=0$ and using the fact that the Coulomb potential is a leaner one $A_0=-ax$,we get $ E=a$. 

\begin{figure}
  \centering
  \includegraphics[scale=1.4]{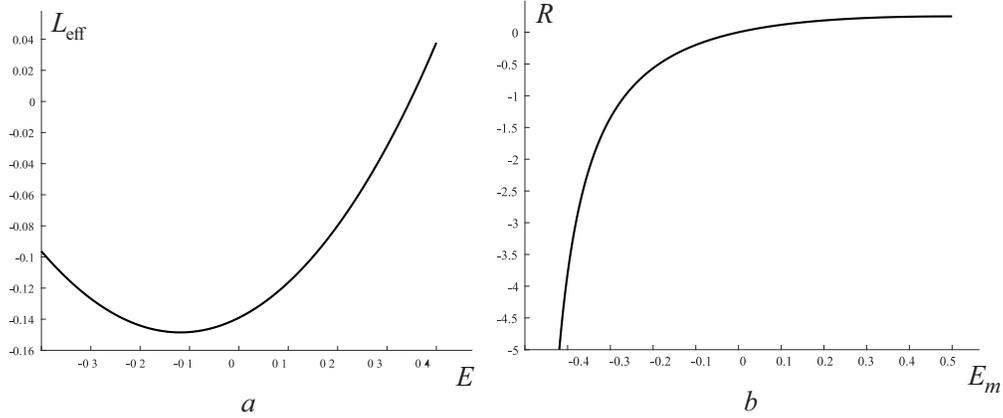}\\
  \caption{a -- The value $L_{\text{eff}}(\phi)$; b -- Chiral condensate $R=\left\langle0|g\bar{\psi}^a\psi^a|0\right\rangle_R$ at $e_0=1$, $m=0,5$ MeV, $M_0^2=100\text{ MeV}^2$.}\label{Ris_4}
\end{figure}

In Fig.~\ref{Ris_4} the effective Lagrangian and the chiral codensate are shown as the function of $E$.

\section{Conclusions}

The models analyzed in this paper, formulated in two dimensional spacetime, are unrealistic. However, we believe that obtained results are showing the method calculation of the chiral condensate in more realistic models.

It was shown that the massive fermions in the massive scalar field have the effective potential with two different vacuums. One of them is global the other is local minima. The chiral condensate  obtained in the model does not disappear if the fermion mass tends to zero. 

The trace of the energy-momentum tensor taking into account the mass of fermions demonstrates a clear violation of discrete symmetry. This fact allowed to construct the effective Lagrangian in QED2. The chiral condensate was obtained in the model. If the fermion mass vanishes, then the chiral condensate disappears.

Technically, the central point is the construction of the effective Lagrangian in QED2.

Acknowledgment: I am grateful to O.V. Kancheli for useful discussions.

\end{document}